\documentclass{article}
\usepackage{ace,amsmath,graphicx,url,times, cite}
\usepackage{graphicx, siunitx}
\usepackage[nolist]{acronym}
\usepackage{amssymb, amsmath, epsfig, rotating, multirow, url}
\usepackage{mathabx} 
\usepackage{epstopdf}
\usepackage{url}
\usepackage{xspace}
\usepackage{hyperref}
\usepackage{balance} 
\pdfoutput=1
\title{Direct-to-Reverberant ratio estimation on the ACE corpus using a Two-channel beamformer}

\name{James Eaton and Patrick A. Naylor}
\address{Department of Electrical and Electronic Engineering, Imperial College London, UK}
\newif\ifarXiv
\arXivfalse
\arXivtrue
\ifarXiv
\newcommand{\expect}{E}
\newcommand{\figsMidSent}{Figs.\@\xspace}
\newcommand{\etal}{\emph{et al.}\@\xspace}
\newcommand{\gHertz}[1]{\SI{#1}{\giga\hertz}}
\newcommand{\dBel}[1]{\SI{#1}{\deci\bel}}
\newcommand{\ExModSq}[1]{\expect\{|{#1}|^{2}\}}
\newcommand{\timeIndex}{t}
\newcommand{\timeDep}{(\timeIndex)}
\newcommand{\sampleIndex}{n}
\newcommand{\sampleDep}{(\sampleIndex)}
\newcommand{\freq}{\omega}
\newcommand{\freqDep}{(j\freq)}
\newcommand{\speech}{x}
\newcommand{\speechT}{\speech\timeDep}
\newcommand{\speechN}{\speech\sampleDep}
\newcommand{\noise}{\nu}
\newcommand{\noiseN}{\noise\sampleDep}
\newcommand{\noiseSpec}{V}
\newcommand{\noisySpeech}{y}
\newcommand{\reverb}{r}
\newcommand{\noisyRevSpeech}{y}
\newcommand{\noisyRevSpeechN}{\noisyRevSpeech\sampleDep}
\newcommand{\noisyRevSpeechSpec}{Y}
\newcommand{\direct}{d}
\newcommand{\convolvedWith}{\ast}
\newcommand{\numMics}{M}
\newcommand{\micIndex}{m}
\newcommand{\DRRPowerRatio}{\eta}
\newcommand{\DRRPowerRatioMean}{\widebar{\DRRPowerRatio}}
\newcommand{\SRRPowerRatio}{\gamma}
\newcommand{\RIR}{h}
\newcommand{\RIRN}{\RIR\sampleDep}
\newcommand{\hmt}{\RIR_{\micIndex}\timeDep}
\newcommand{\hdmt}{\RIR_{\micIndex,\direct}\timeDep}
\newcommand{\hrmt}{\RIR_{\micIndex,\reverb}\timeDep}
\newcommand{\ymt}{\noisyRevSpeech_{\micIndex}\timeDep}
\newcommand{\mth}{$\micIndex$-th}
\newcommand{\vmt}{\noise_{\micIndex}\timeDep}
\newcommand{\DSBOutputSpec}{Z}
\newcommand{\BFGainSpec}{G}
\newcommand{\fLow}{\freq_{1}}
\newcommand{\fHigh}{{\freq_{2}}}
\else

\fi
\newif\ifshowSubbands
\showSubbandstrue
\newcommand{\Falk}{A}
\newcommand{\Jeub}{B}
\newcommand{\EatonNoNR}{C}
\newcommand{\EatonSpecSub}{D}
\newcommand{\EatonGerkmann}{E}
\newcommand{\EatonFFT}{F}
\newcommand{\EatonFilt}{G}
\begin{document}
\begin{acronym}
\ifarXiv
\acro{ACE}{Acoustic Characterization of Environments\acroextra{. An IEEE challenge run by the SAP group at Imperial College}}
\acro{AI}{Articulation Index}
\acro{AIR}{Acoustic Impulse Response}
\acro{DENBE}{\ac{DRR} Estimation using a Null-Steered Beamformer}
\acro{DoA}{Direction-of-Arrival}
\acro{DOA}{Direction-of-Arrival}
\acro{DR}{Douglas-Rachford}
\acro{DRR}{Direct-to-Reverberant Ratio}
\acro{ISO}{International Organization for Standardization}
\acro{MMSE}{Minimum Mean Squared Error}
\acro{RIR}{Room Impulse Response}
\acro{RT}{Reverberation Time}
\acro{RTF}{Real-Time Factor}
\acro{SNR}{Signal-to-Noise Ratio}
\acro{SR}{Sparse Representation}
\acro{SRR}{Signal-to-Reverberation Ratio}
\acro{STFT}{Short Time Fourier Transform}
\acro{T60}[$T_\textrm{60}$]{Reverberation Time\acroextra{ to decay by $60$ dB}}
\acro{TI}{Texas Instruments, Inc.}
\acro{TIMIT}{\ac{TI}-\ac{MIT} speech corpus}

\else
\input{../SapBibTex/sapacronyms.txt}
\fi
\end{acronym}
\ninept
\maketitle
\begin{sloppy}
\begin{abstract}
\ac{DRR} is an important measure for characterizing the properties of a room.
The recently proposed DRR Estimation using a Null-Steered Beamformer (DENBE) algorithm was originally tested on simulated data where noise was artificially added to the speech after convolution with impulse responses simulated using the image-source method.
This paper evaluates the performance of this algorithm on speech convolved with measured impulse responses and noise using the \ac{ACE} Evaluation corpus.
The fullband \ac{DRR} estimation performance of the DENBE algorithm exceeds that of the baselines in all \acp{SNR} and noise types.
\ifshowSubbands
In addition, estimation of the \ac{DRR} in one third-octave \acs{ISO} frequency bands is demonstrated.
\fi
\end{abstract}

\acresetall
\section{Introduction}
\label{sec:intro}
The acoustic properties of a room can be described by its \ac{AIR}.
From the \ac{AIR}, measured for example using an exponential sinusoidal sweep~\cite{Farina2000}, the acoustic parameters of \ac{T60} and \ac{DRR} can be determined using standardized methods such as~\cite{ISO_3382}.
The performance of speech enhancement and speech recognition may be assisted with knowledge of these parameters.
In practical situations such as mobile telecommunications, though, the \ac{AIR} is seldom available, or easy to estimate.
In which case the acoustic parameters describing the room must be determined from the speech recorded in the room using blind estimators such as~\cite{Eaton2013}.

The presence of ambient and sensor noise must also be considered.
%
The total signal model is as follows:
\begin{equation}
\noisyRevSpeechN = \speechN \convolvedWith \RIRN + \noiseN, 
\label{eqn:noisySpeech}
\end{equation}
%
%
where $ \noisyRevSpeechN$ is the sampled noisy reverberant speech, $\speechN$ is the sampled anechoic speech, $\RIRN$ is the sampled \ac{AIR}, and $\noiseN$ is the sampled additive noise.

It is common practice in the field of speech processing to use simulated noisy reverberant speech based on anechoic speech corpi such as \acs{TIMIT}~\cite{Garofolo1988} convolved with simulated \acp{AIR} using the image-source method~\cite{Allen1979}, and with additive noise from corpi such as NOISEX-92 \cite{Varga1993}.
A drawback of using pre-recorded noise sources is that any additive noise may have been recorded in a substantially different environment to the \ac{AIR} being simulated.
%
Without the \acp{AIR} of the noises sources, pre-recorded noise may lead to misleading estimation results when estimating non-intrusively from such noisy reverberant speech because the decays identified by the estimator may originate from the acoustic environment of the pre-recorded noise at low \acp{SNR}.
A commonly used source of babble noise~\cite{Varga1993} has a clearly audible room acoustic.

%
Part of the motivation for the \ac{ACE} corpus~\cite{Eaton2015a} was to overcome this problem.
The \ac{ACE} corpus provides a set of matched recorded \acp{AIR} and noises including ambient, fan and babble noise.
The babble noise was recorded using multiple speakers recorded in the same room as the \ac{AIR} under the same conditions.

Whilst the \ac{DENBE} algorithm~\cite{Eaton2015} was tested in simulated data, the \ac{ACE} challenge has provided the opportunity to evaluate the algorithm in more realistic scenarios where the \ac{DoA} of the speech is not known and must be accommodated by the algorithm, and also the noise must be estimated.
The \ac{DRR} estimate will be affected differently depending on whether the noise is coherent, such as non-stationary noise, or diffuse.
\ifshowSubbands
In addition, the \ac{ACE} corpus provides ground truth \ac{DRR} in \acs{ISO} frequency bands~\cite{ISO_266}.  
The capability of the \ac{DENBE} algorithm to estimate \ac{DRR} in \acs{ISO} frequency bands is therefore explored using two different approaches based on filtering before the algorithm, and transforming the frequency dependent \ac{DRR} available internally within the algorithm to the \acs{ISO} preferred frequencies.

A particular challenge in estimating in frequency bands is that the bandwidth of speech does not occupy all of the \acs{ISO} frequencies up to half the sample rate, and where there is no speech, noise will dominate.
Estimating \ac{DRR} in \acs{ISO} frequency bands non-intrusively from speech is only being attempted very recently, and at the time of writing there are no published algorithmic papers on the subject.
\fi

The contribution of this paper is therefore to evaluate the recently proposed~\ac{DENBE} algorithm~\cite{Eaton2015} on the \ac{ACE} corpus alongside the methods of Falk {\etal}~\cite{Falk2009}, and Jeub {\etal}~\cite{Jeub2011}
\ifshowSubbands
both in fullband and in one third-octave \acs{ISO} subbands.
\else
in fullband.
\fi

The remainder of the paper is organized as follows:
In Section~\ref{sec:review}, the \ac{DENBE} algorithm is reviewed.
In Section~\ref{sec:perfeval}, the approach to the performance evaluation is described.
In Section~\ref{sec:results}, the results of the experiments
\ifshowSubbands
including the results of \ac{DRR} estimation in \acs{ISO} frequency bands
\fi
are discussed and in Section~\ref{sec:conc}, conclusions are drawn.

\section{Review of the \ac{DENBE} Algorithm}
\label{sec:review}
A speech signal, $\speech\timeDep$, continuous in time, radiating from a position in a room will follow multiple paths to any observation point comprising the direct path as well as reflections from the surfaces and objects in the room.
The reverberant signal at the {\mth} microphone in an array of $\numMics$ microphones $\noisyRevSpeech_{\micIndex}\timeDep$, in the room, is characterised by $\RIR_{\micIndex}\timeDep$, the \ac{AIR} of the acoustic channel between the source and the {\mth} microphone as
\begin{equation}
\ymt = \hmt \convolvedWith  \speechT  + \vmt,
\label{eqn:SigAtMicmt}
\end{equation}
where $\vmt$ is the additive noise at the {\mth} microphone.
The \ac{AIR} is a function of the room geometry, absorption of the surfaces of the room and its contents, the source and the microphone locations.
Let
\begin{equation}
\hmt = \hdmt + \hrmt,
\label{eqn:AIR}
\end{equation}
where $\hdmt$ and $\hrmt$ are the impulse responses of the direct and reverberant paths for the $\micIndex$-th microphone respectively.
The \ac{DRR} at the {\mth} microphone, $\DRRPowerRatioMean_{\micIndex}$, is the ratio of the power arriving directly at the microphone from the source to power arriving after being reflected from one or more surfaces in the room~\cite{Naylor2010b}.
It can be written as
%
%
\begin{equation}
\DRRPowerRatioMean_{\micIndex} =  
\frac{
\int |\RIR_{d,\micIndex}\timeDep|^2 d\timeIndex
}{
\int |\RIR_{r,\micIndex}\timeDep|^2 d\timeIndex
}.
\label{eqn:DRR}
\end{equation}
When the impulse response is convolved with a speech signal, the observation at the {\mth} microphone is the \ac{SRR}, $\SRRPowerRatio$, given by
\begin{equation}
\SRRPowerRatio_{\micIndex} =
\frac{
\ExModSq{(\RIR_{d,\micIndex}\timeDep)^{T}  \convolvedWith \speech\timeDep}  
}{
\ExModSq{(\RIR_{r,\micIndex}\timeDep)^{T}  \convolvedWith \speech\timeDep} 
}.
\label{eqn:DRRSpec}
\end{equation}
The \ac{SRR} is equal to the \ac{DRR} in the case when $\speech\timeDep$ is spectrally white.

As the length of the signal $\speech\timeDep$ increases, ${\SRRPowerRatio_{\micIndex}}$ at a given frequency will approach the \ac{DRR}.
In the \ac{DENBE} algorithm, a null-steered beamformer is used to obtain the reverberant signal which is then rendered in \ac{STFT} frequency bands.
The null in the beamformer is steered at the source thus cancelling the signal emanating directly from the source in the output of the beamformer.
The output of the beamformer will thus consist largely of the reverberant signal.
The amount of cancellation of the source is dependent on the frequency and the beamformer weights.
Since the total signal is known, and the gain of the beamformer can be computed at a given frequency, it is therefore possible to determine the direct signal.
After some manipulation then
\begin{equation}
\DRRPowerRatio_{\micIndex}\freqDep 
=
\frac{
\ExModSq{\noisyRevSpeechSpec_{\micIndex}\freqDep}
-
\ExModSq{\noiseSpec_{\micIndex}\freqDep}
}{
\frac{1}{\BFGainSpec^2\freqDep}(
\ExModSq{\DSBOutputSpec_{\noisySpeech}\freqDep}  
-
\ExModSq{\DSBOutputSpec_{\noise}\freqDep})
}
 - 1.
\label{eqn:DRRPowerRatioCalc}
\end{equation}
where $\noisyRevSpeechSpec_{\micIndex}\freqDep$ is the total noisy reverberant speech spectrum at the $\micIndex$th microphone,
$\noiseSpec_{\micIndex}\freqDep$ is the total noise spectrum at the $\micIndex$th microphone,
$\BFGainSpec\freqDep$ is the frequency dependent gain of the beamformer, 
$\DSBOutputSpec_{\noisySpeech}\freqDep$ is the total beamformer output,
and 
$\DSBOutputSpec_{\noise}\freqDep$ is the noise at the beamformer output.
The overall \ac{DRR} is then given by
\begin{equation}
\DRRPowerRatioMean_{\micIndex} = \frac{1}{\fHigh-\fLow} \int_{\fLow}^{\fHigh} \DRRPowerRatio_{\micIndex}\freqDep \text{ } d\freq,
\label{eqn:w1w2}
\end{equation}
where
$\freq_{1} \leq \freq \leq \freq_{2}$ is the frequency range of interest.
The frequency dependent \ac{DRR} in one third-octave \acs{ISO} frequency bands is obtained by determining a matrix of weights for each \ac{STFT} frequency band relative to one third-octave \acs{ISO} frequency bands and then multiplying the vector of frequency dependent \ac{STFT} \acs{DRR} by the weights matrix.
%
\begin{figure}[!hb]
	\ifarXiv
	\centerline{\epsfig{figure=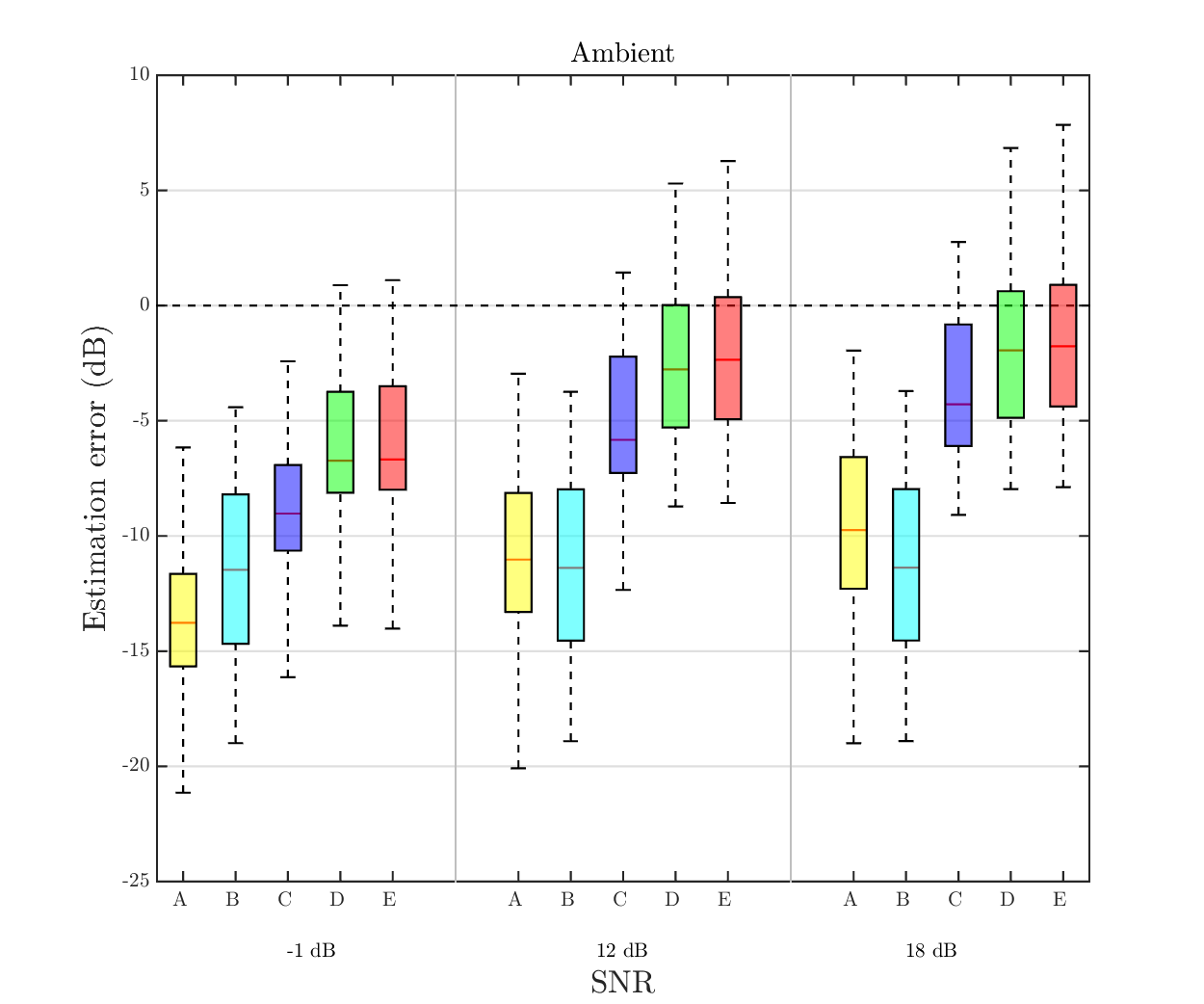,
	width=80mm,viewport=30 5 552 467,clip}}%
	\else
	\centerline{\epsfig{figure=Figs/ana_eval_gt_partic_results_combined_Phase3_Eaton_P3_DRR_dB_by_SNR_Ambient.png,
	width=80mm,viewport=30 5 552 467,clip}}%
	\fi
	\caption{{\ac{DRR} estimation error for Ambient noise for computation methods {\Falk} to {\EatonGerkmann} for low (\dBel{-1}), medium (\dBel{12}) and high (\dBel{18}) \acp{SNR}}}%
\label{fig:ACE_Ambient}%
\end{figure}%
\begin{figure}[!ht]
	\ifarXiv
	\centerline{\epsfig{figure=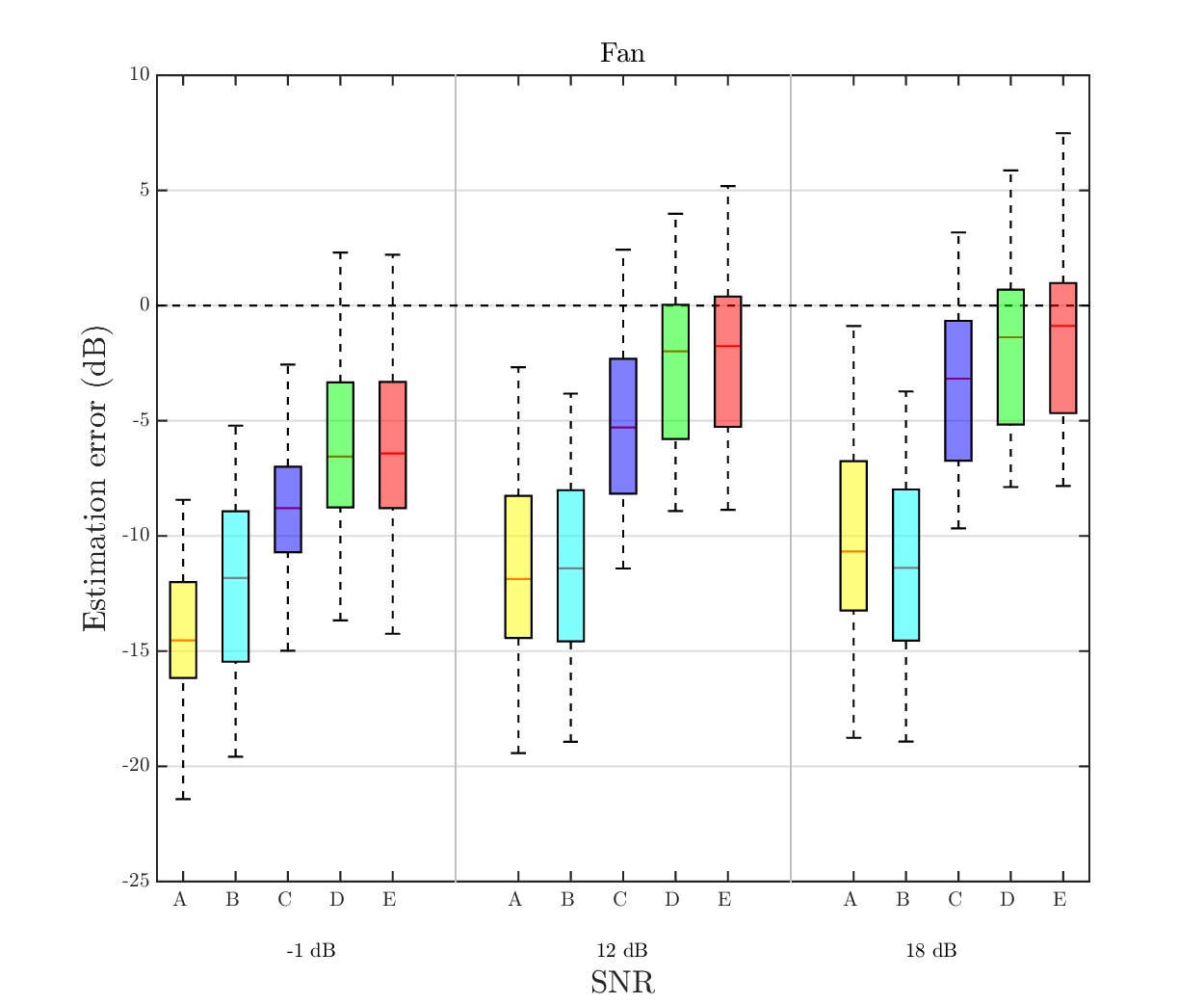,
	width=80mm,viewport=30 5 552 467,clip}}%
	\else
	\centerline{\epsfig{figure=Figs/ana_eval_gt_partic_results_combined_Phase3_Eaton_P3_DRR_dB_by_SNR_Fan.png,
	width=80mm,viewport=30 5 552 467,clip}}%
	\fi
	\caption{{\ac{DRR} estimation error for Fan noise for computation methods {\Falk} to {\EatonGerkmann} for low (\dBel{-1}), medium (\dBel{12}) and high (\dBel{18}) \acp{SNR}}}%
\label{fig:ACE_Fan}%
\end{figure}%
\begin{figure}[!ht]
	\ifarXiv
	\centerline{\epsfig{figure=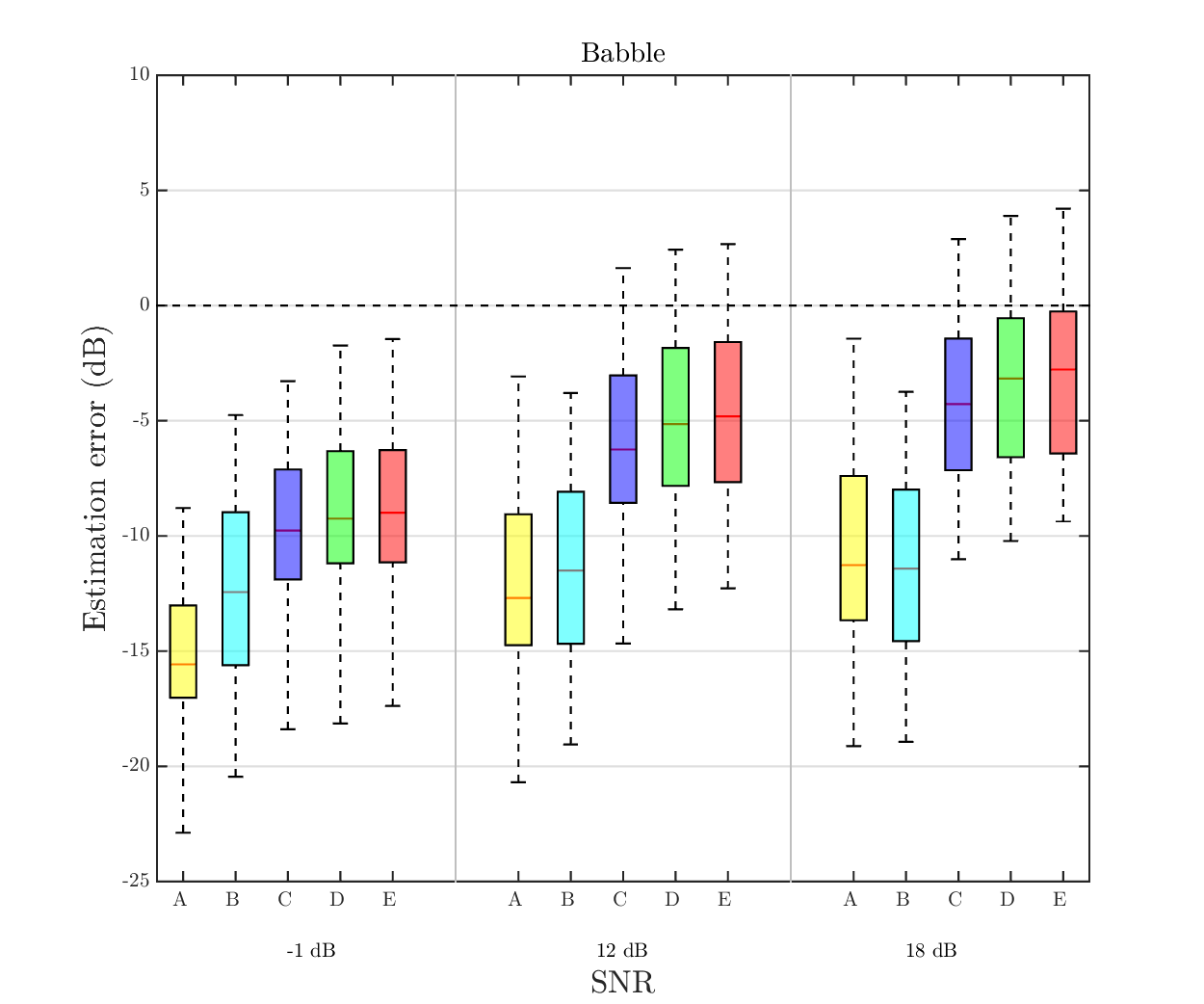,
	width=80mm,viewport=30 5 552 467,clip}}%
	\else
	\centerline{\epsfig{figure=Figs/ana_eval_gt_partic_results_combined_Phase3_Eaton_P3_DRR_dB_by_SNR_Babble.png,
	width=80mm,viewport=30 5 552 467,clip}}%
	\fi
	\caption{{\ac{DRR} estimation error for Babble noise for computation methods {\Falk} to {\EatonGerkmann} for low (\dBel{-1}), medium (\dBel{12}) and high (\dBel{18}) \acp{SNR}}}%
\label{fig:ACE_Babble}%
\end{figure}%
\ifshowSubbands
\begin{figure}[!ht]
	\ifarXiv
	\centerline{\epsfig{figure=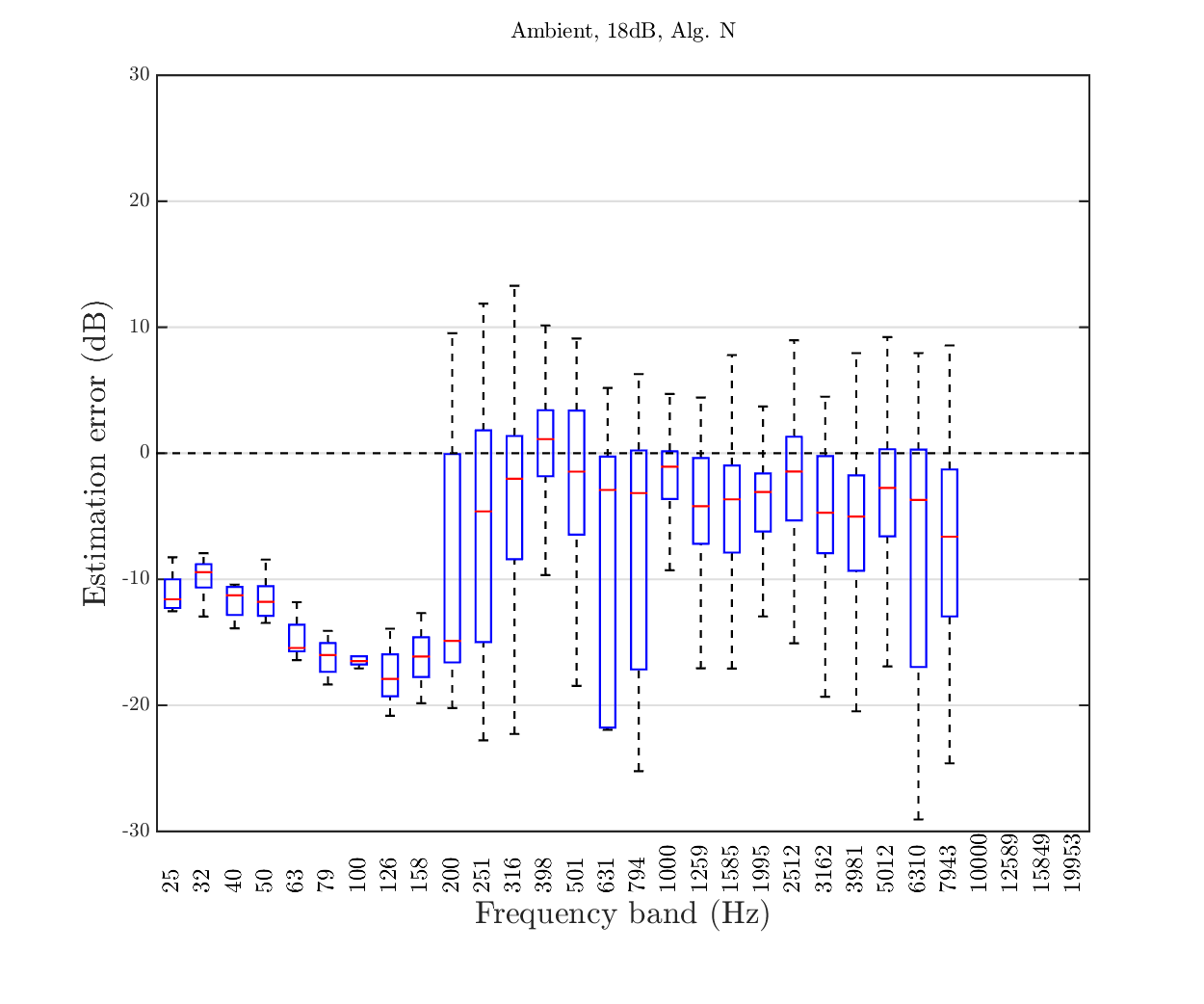,
	width=80mm,viewport=30 37 552 475,clip}}%
	\centerline{\epsfig{figure=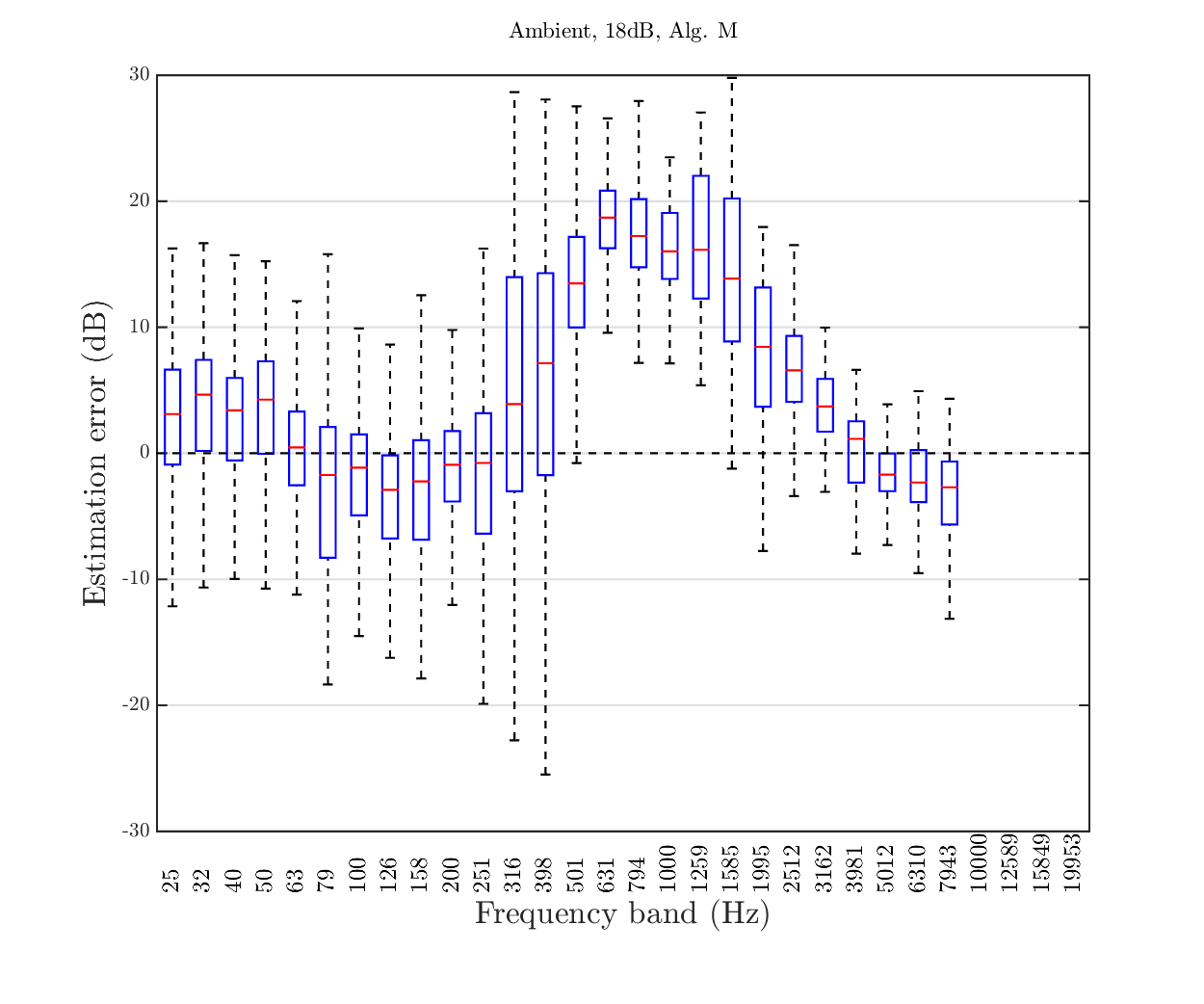,
	width=80mm,viewport=30 37 552 475,clip}}%
	\else
	\centerline{\epsfig{figure=Figs/ana_eval_gt_partic_results_combined_Phase3_Eaton_Thesis_P3S_DRR_dB_18dB_SNR_Ambient_sub_ICASSP_2015_DRR_2-ch_Gerkmann_NR_FFT_subband.png,
	width=80mm,viewport=30 37 552 475,clip}}%
	\centerline{\epsfig{figure=Figs/ana_eval_gt_partic_results_combined_Phase3_Eaton_Thesis_P3S_DRR_dB_18dB_SNR_Ambient_sub_ICASSP_2015_DRR_2-ch_Gerkmann_NR_filtered_subband.png,
	width=80mm,viewport=30 37 552 475,clip}}%
	\fi
	\caption{{\ac{DRR} estimation error for Ambient noise for computation methods {\EatonFFT} (upper plot) and {\EatonFilt} (lower plot) for high (\dBel{18}) \acp{SNR}}}%
\label{fig:ACE_Ambient2}%
\end{figure}%
\begin{figure}[!ht]
	\ifarXiv
	\centerline{\epsfig{figure=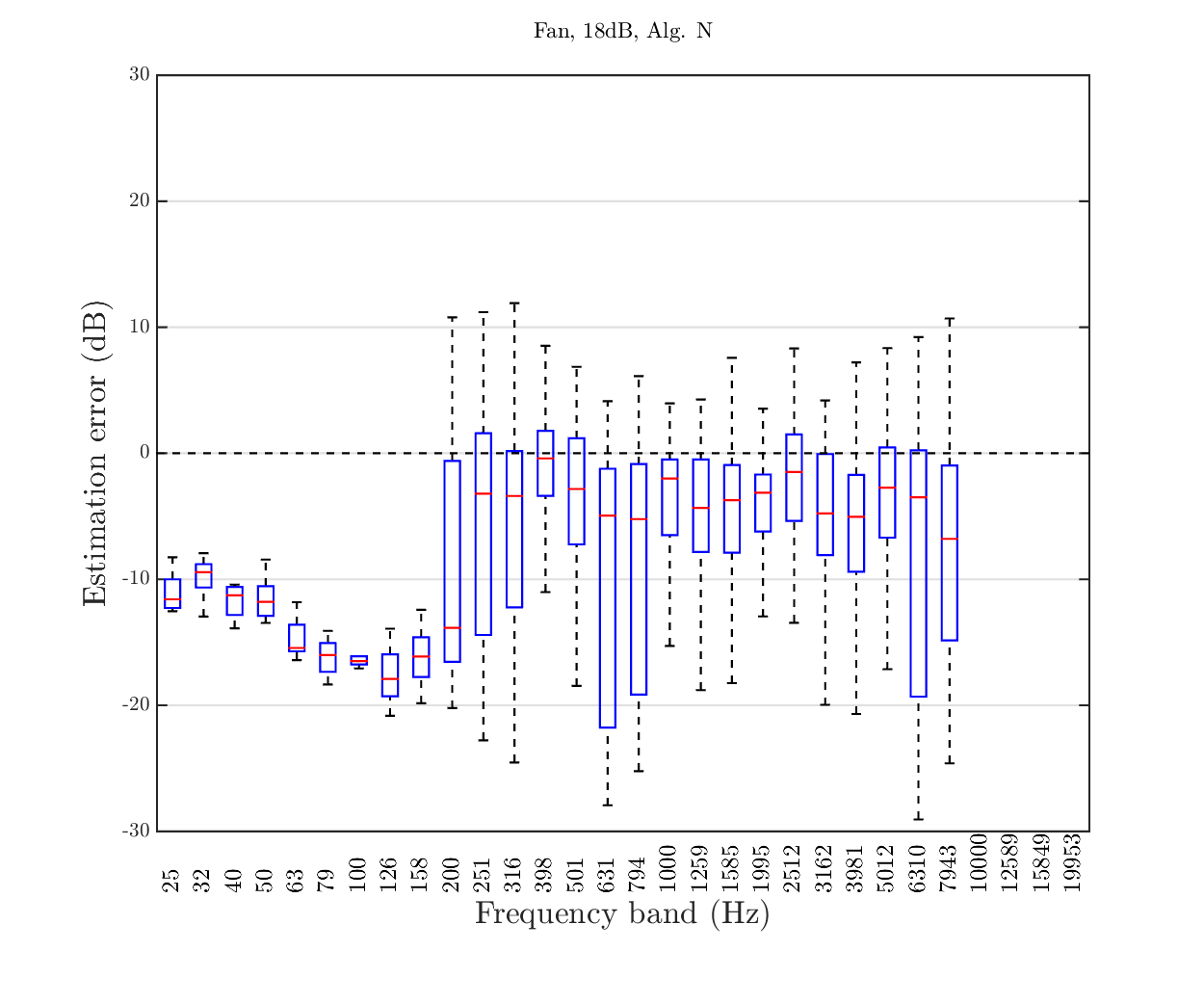,
	width=80mm,viewport=30 37 552 475,clip}}%
	\centerline{\epsfig{figure=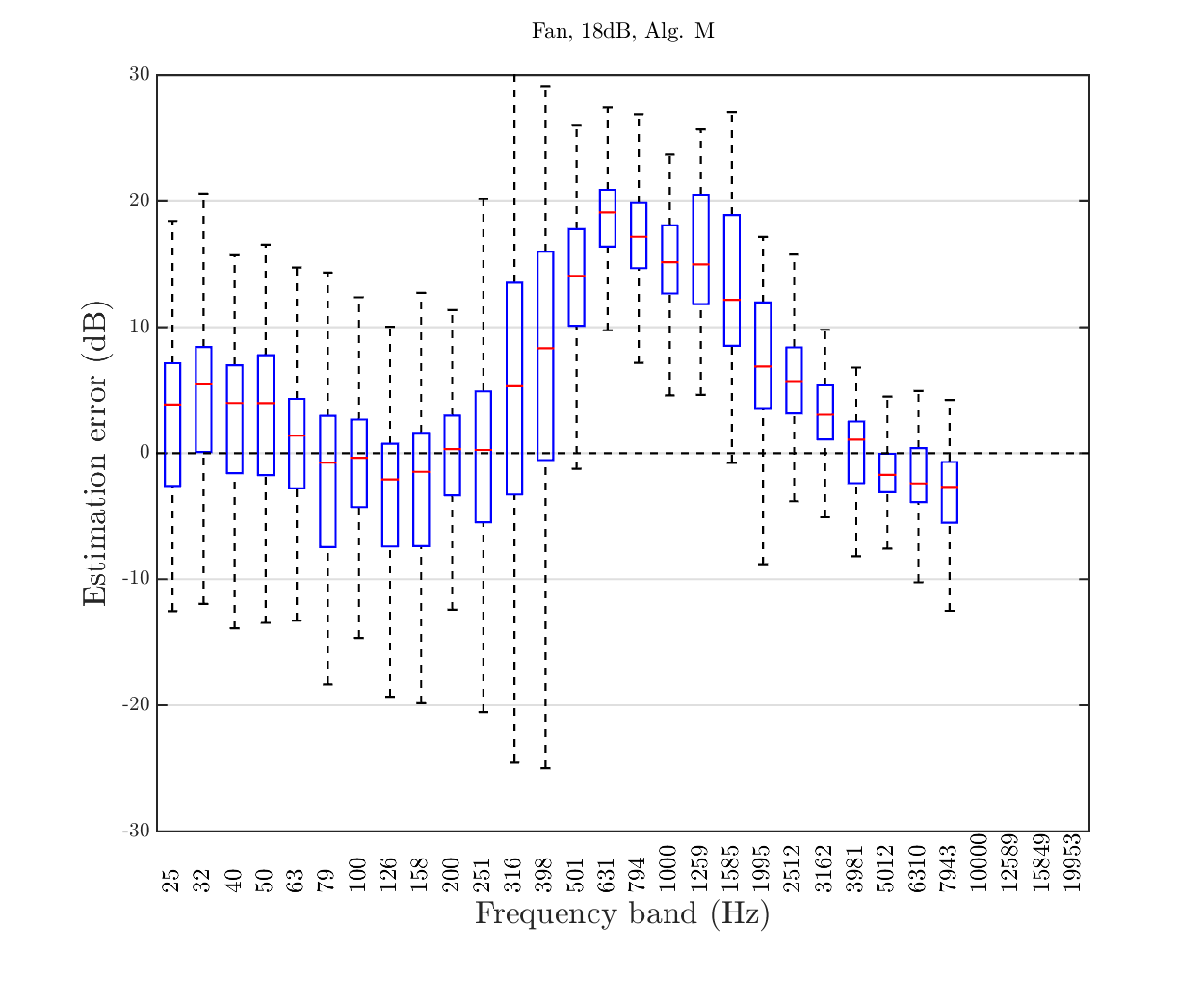,
	width=80mm,viewport=30 37 552 475,clip}}%
	\else
	\centerline{\epsfig{figure=Figs/ana_eval_gt_partic_results_combined_Phase3_Eaton_Thesis_P3S_DRR_dB_18dB_SNR_Fan_sub_ICASSP_2015_DRR_2-ch_Gerkmann_NR_FFT_subband.png,
	width=80mm,viewport=30 37 552 475,clip}}%
	\centerline{\epsfig{figure=Figs/ana_eval_gt_partic_results_combined_Phase3_Eaton_Thesis_P3S_DRR_dB_18dB_SNR_Fan_sub_ICASSP_2015_DRR_2-ch_Gerkmann_NR_filtered_subband.png,
	width=80mm,viewport=30 37 552 475,clip}}%
	\fi
	\caption{{\ac{DRR} estimation error for Fan noise for computation methods {\EatonFFT} (upper plot) and {\EatonFilt} (lower plot) for high (\dBel{18}) \acp{SNR}}}%
\label{fig:ACE_Fan2}%
\end{figure}%
\begin{figure}[!ht]
	\ifarXiv
	\centerline{\epsfig{figure=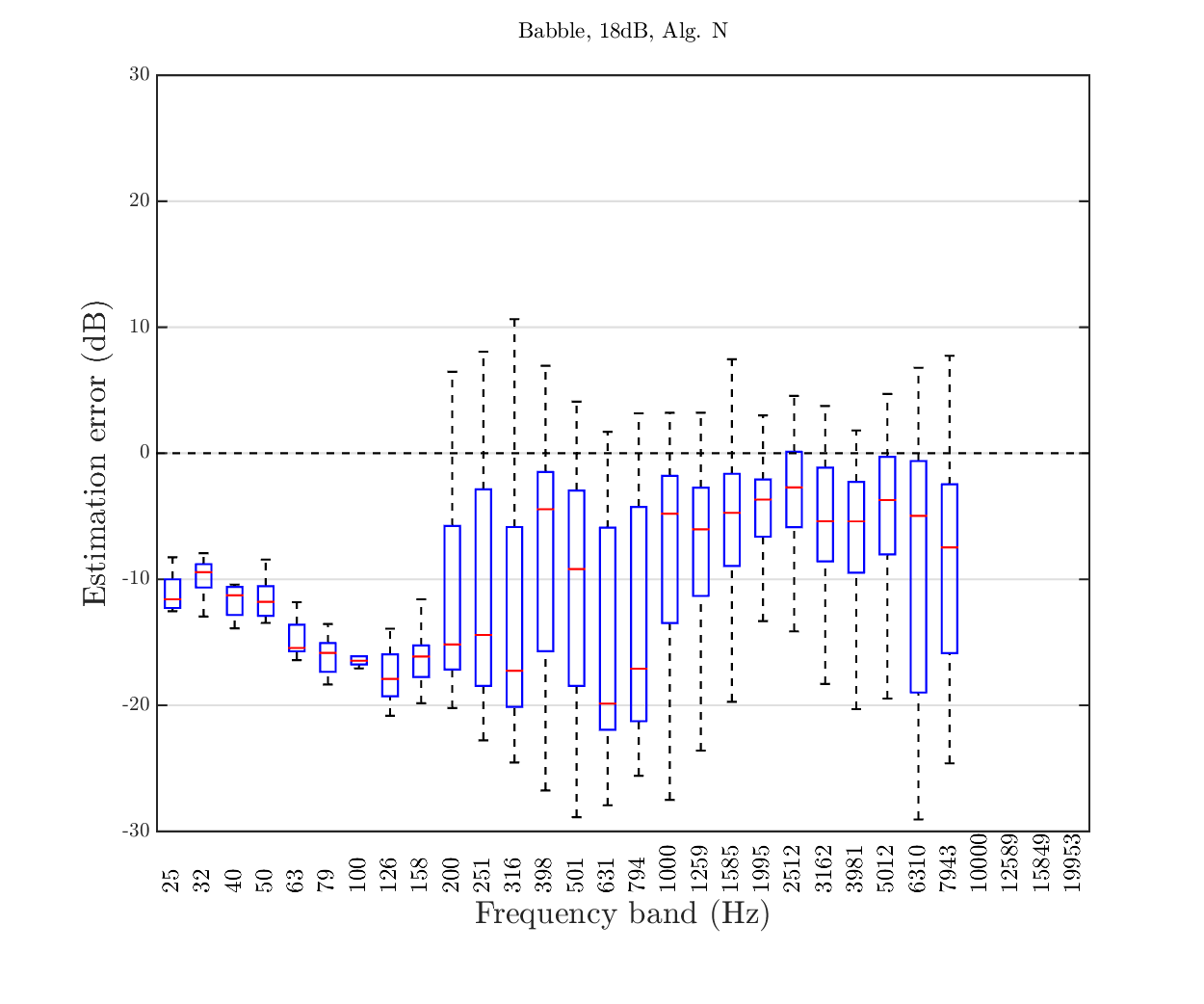,
	width=80mm,viewport=30 37 552 475,clip}}%
	\centerline{\epsfig{figure=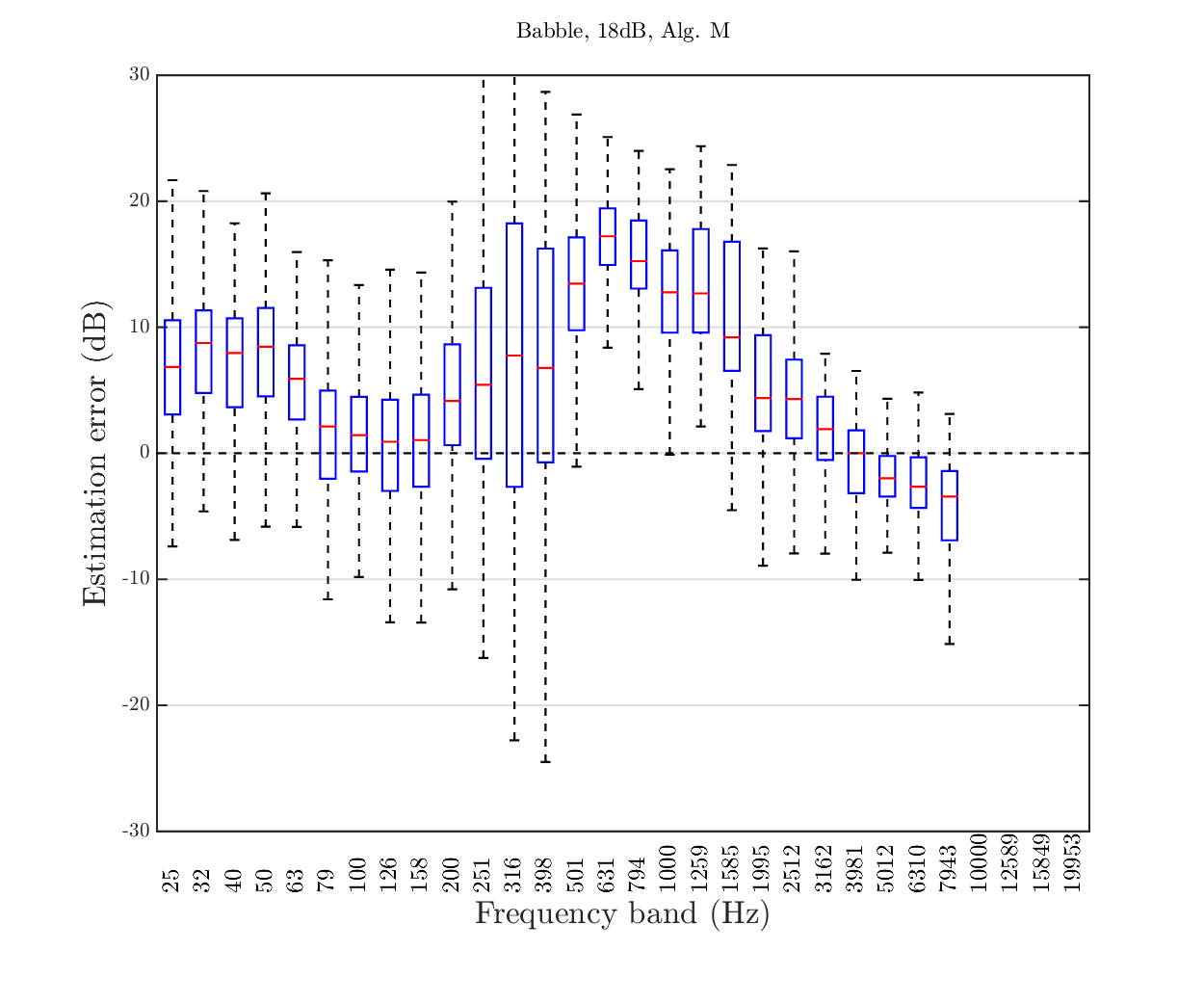,
	width=80mm,viewport=30 37 552 475,clip}}%
	\else
	\centerline{\epsfig{figure=Figs/ana_eval_gt_partic_results_combined_Phase3_Eaton_Thesis_P3S_DRR_dB_18dB_SNR_Babble_sub_ICASSP_2015_DRR_2-ch_Gerkmann_NR_FFT_subband.png,
	width=80mm,viewport=30 37 552 475,clip}}%
	\centerline{\epsfig{figure=Figs/ana_eval_gt_partic_results_combined_Phase3_Eaton_Thesis_P3S_DRR_dB_18dB_SNR_Babble_sub_ICASSP_2015_DRR_2-ch_Gerkmann_NR_filtered_subband.png,
	width=80mm,viewport=30 37 552 475,clip}}%
	\fi
	\caption{{\ac{DRR} estimation error for Babble noise for computation methods {\EatonFFT} (upper plot) and {\EatonFilt} (lower plot) for high (\dBel{18}) \acp{SNR}}}%
\label{fig:ACE_Babble2}%
\end{figure}%
\fi
%
\section{Performance evaluation}
\label{sec:perfeval}
The performance evaluation of the \ac{DENBE} algorithm was performed using the Chromebook (laptop) two-channel dataset of the Evaluation stage software provided by the \ac{ACE} challenge.
The \ac{ACE} challenge corpus~\cite{Eaton2015a} comprises \num{4500} noisy reverberant speech files.
This is based on 5 male and 5 female talkers with 5 utterances each of different lengths of anechoic speech.
Three different \acp{SNR} are used: High (\dBel{18}), Medium (\dBel{12}), and Low (\dBel{-1}). 
Three different noise types are applied: Ambient which is the sound of the room with no speech, Fan, the sound of the room with one or more fans operating, and Babble, the sound of multiple talkers speaking simultaneously in the room reading from \acs*{TIMIT} passages or scientific papers.
\acp{AIR} from 5 different rooms each with two different microphone positions are convolved with the anechoic speech and then mixed with noise using the \emph{v\_addnoise} function~\cite{Brookes1997}.

The \ac{DENBE} algorithm requires the ground truth noise and the \ac{DoA}.
Since the ground truth noise and \ac{DoA} were not available, additional pre-processing was performed resulting in several variants of the \ac{DRR} estimation algorithm test.
In all variants time alignment of the two input channels was applied using \emph{sigalign.m}~\cite{Brookes1997}.
In the first variant denoted {\EatonNoNR}, no noise reduction was applied, and fullband \ac{DRR} was estimated.
In the second variant, denoted {\EatonSpecSub}, spectral subtraction was performed using minimum statistics~\cite{Martin2001,Brookes1997}, and fullband \ac{DRR} estimated.
In the third variant, denoted {\EatonGerkmann}, spectral subtraction was performed using \ac{MMSE} power estimation~\cite{Gerkmann2012,Brookes1997}, and fullband \ac{DRR} estimated.
\ifshowSubbands

A further two variants of third variant were devised.
The third variant was chosen for further experiments since it was shown in~\cite{Eaton2013b} that the \ac{MMSE} noise power estimation was the most effective in the \ac{SNR} range of the \ac{ACE} Challenge datasets.
In the fifth variant, denoted {\EatonFFT}, \ac{DRR} estimates in one third-octave \acs{ISO} frequency bands were derived by transforming the results of the \ac{DRR} estimate for each \ac{STFT} frequency band in \eqref{eqn:DRRPowerRatioCalc} into one third-octave \acs{ISO} frequency bands.
In the sixth variant, denoted {\EatonFilt}, \ac{DRR} estimates in one third-octave \acs{ISO} frequency bands were obtained by filtering the input to the algorithm using an 8th order Butterworth filter and estimating the \ac{DRR} using \eqref{eqn:w1w2} over all \ac{STFT}frequency bands as in the fullband \ac{DENBE} algorithm.
\fi
In addition, two baselines were used for comparison of the fullband \ac{DRR} results,
Falk {\etal}~\cite{Falk2009}, 
and
Jeub {\etal}~\cite{Jeub2011}
denoted {\Falk} and {\Jeub} respectively.  
\ifshowSubbands
Whilst the \ac{DENBE} algorithm is computationally efficient, it is expected that variant {\EatonFilt} will have a much higher \ac{RTF} than the other variants of \ac{DENBE} since the fullband \ac{DRR} estimation must be performed in every frequency band up to half the sample rate.
\fi

In addition to estimation performance, the \ac{RTF} of each method was compared.
Algorithm {\Falk} was tested using Matlab on an Intel Xeon X5675 processor with a clock speed of \gHertz{3.07},
whilst algorithms {\Jeub} to {\EatonFilt} were tested using Matlab on an Intel Xeon E5-2643 processor with a clock speed of \gHertz{3.30}.  
These two processors have similar levels of performance.
%
\section{Results}
\label{sec:results}
Results for algorithms {\Falk} to {\EatonGerkmann} in the three noise types, Ambient, Fan, and Babble are shown in {\figsMidSent}~\ref{fig:ACE_Ambient},~\ref{fig:ACE_Fan}, and~\ref{fig:ACE_Babble} respectively.
On each box, the central mark is the median, the edges of the box are the \num{25}th and \num{75}th percentiles,
the whiskers extend to the most extreme data points not considered outliers.
Outliers are not shown.
Method {\EatonNoNR} gives better performance than the baselines under all conditions.
With the addition of noise reduction in methods {\EatonSpecSub} and {\EatonGerkmann}, the estimation performance is greatly improved for fan and ambient noise regardless of the method of noise reduction.

\ifshowSubbands
Results in one third-octave \acs{ISO} frequency bands for algorithms {\EatonFFT} and {\EatonFilt} in the three noise types at \dBel{18} \acp{SNR} are shown in {\figsMidSent} \ref{fig:ACE_Ambient2} to \ref{fig:ACE_Babble2}.
Algorithm {\EatonFFT} achieves some success at \SI{200}{\hertz} and above in ambient and fan noise, and above \SI{1200}{\hertz} in babble noise.
Below \SI{200}{\hertz} the algorithm is is unable to estimate the \ac{DRR} and is returning a value of \dBel{-20} in most cases.
As such the error is large and negative with a small variance.
Algorithm {\EatonFilt} achieves better performance below around \SI{250}{\hertz}.
This suggests that a combination of the two algorithms using the results of {\EatonFFT} above \SI{200}{\hertz} and the results of {\EatonFilt} below \SI{200}{\hertz} would be a better estimator than either estimator on its own.
However, the variance in most frequency bands is large.
Further investigation is required to determine whether this information would be useful to a dereverberation algorithm or a speech recognizer.
\fi

Table~\ref{tab:performance} shows the \ac{RTF} for each algorithm computed by dividing the total CPU time used to process all \num{4500} noisy reverberant speech files in the \ac{ACE} Evaluation dataset by the combined length of all the noisy reverberant speech files.
\begin{table} [!htb] 
\caption{\label{tab:performance} {Comparison of \ac{RTF}}} 
\vspace{2mm} 
\centerline{
\ifshowSubbands 
\begin{tabular}{|c|c|c|c|c|c|c|} 
\else
\begin{tabular}{|c|c|c|c|c|} 
\fi
\hline 
\ifshowSubbands
{\Falk}~\cite{Falk2009}   
&{\Jeub}~\cite{Jeub2011}   
&{\EatonNoNR}~\cite{Eaton2015} 
&{\EatonSpecSub}~\cite{Eaton2015} 
&{\EatonGerkmann}~\cite{Eaton2015}
&{\EatonFFT}~\cite{Eaton2015}
&{\EatonFilt}~\cite{Eaton2015} \\ 
\else
{\Falk}~\cite{Falk2009}   
&{\Jeub}~\cite{Jeub2011}   
&{\EatonNoNR}~\cite{Eaton2015} 
&{\EatonSpecSub}~\cite{Eaton2015} 
&{\EatonGerkmann}~\cite{Eaton2015} \\ 
\fi
\hline	\hline 
\ifshowSubbands
\num{0.082}
&\num{0.016}    
&\num{0.029}  
&\num{0.054}       
&\num{0.038}       
&\num{0.036}       
&\num{0.69}     \\ 
\else
\num{0.082}
&\num{0.016}    
&\num{0.029}  
&\num{0.054}       
&\num{0.038}     \\ 
\fi
	\hline 
\end{tabular}}
\end{table}
\section{Conclusion}
\label{sec:conc}
The \ac{ACE} Challenge has provided an opportunity to evaluate the \ac{DENBE} algorithm on real recordings of \acp{AIR} and noise in contrast to the previously reported performance which was on simulated data.
Also, the speech in the \ac{ACE} corpus is free-running and less uniform than the \acs{TIMIT} database used in~\cite{Eaton2015}, and so is representative of natural rather than read speech.
The estimation performance in fullband compares very favourably with existing methods, and has a low \ac{RTF} making it suitable for real-time applications.
\ifshowSubbands
The performance in one third-octave \acs{ISO} frequency bands has also been demonstrated.
This is a very recent area of research and the initial results based on small additions to the published algorithm produce encouraging results.
\fi
\clearpage
\balance
\bibliographystyle{IEEEtran}
\bibliography{../SapBibTex/sapref}
\label{sec:bibliography}
\end{sloppy}
\end{document}